\documentclass[aps,prd,twocolumn,superscriptaddress,showpacs]{revtex4}
\usepackage{dcolumn,natbib,multirow} 
\usepackage{graphicx,amsmath}
\usepackage{bm}

\newcommand{\hMsun}{h^{-1}M_{\odot}}

\newcommand{\sun}{\ensuremath{\odot}}

\newcommand{\ds}{\ensuremath{\Delta\Sigma}}
\newcommand{\hmpc}{$h^{-1}$Mpc}
\newcommand{\hkpc}{$h^{-1}$kpc}
\newcommand{\hmsun}{\ensuremath{h^{-1}M_{\odot}}}

\newcommand{\beq}{\begin{equation}}
\newcommand{\eeq}{\end{equation}}
\newcommand{\beqa}{\begin{eqnarray}}
\newcommand{\eeqa}{\end{eqnarray}}

\begin{document}

\title{Halo mass - concentration relation from weak lensing}

\author{Rachel Mandelbaum}\email{rmandelb@ias.edu},\thanks{Hubble fellow}
\affiliation{Institute for Advanced Study, Einstein Drive, Princeton NJ
  08540, USA}
\author{Uro\v s Seljak}
\affiliation{Institute for Theoretical Physics, University of Zurich, Zurich Switzerland}
\affiliation{Department of Physics, University of California, Berkeley, CA 94720, USA} 
\author{Christopher M. Hirata}
\affiliation{Mail Code 130-33, Caltech, Pasadena, CA 91125, USA}

\date{\today}

\begin{abstract}
We perform a statistical weak lensing analysis of dark matter profiles around 
tracers of halo mass from galactic- to cluster-size halos. In this analysis we use
170~640 isolated $\sim L_*$ galaxies split into 
ellipticals and spirals, 38~236 groups traced by isolated spectroscopic 
Luminous Red Galaxies (LRGs) and 13~823 MaxBCG clusters 
from the Sloan Digital Sky Survey (SDSS) covering a wide range of richness. Together these three samples allow
a determination of the density profiles of dark matter halos over three orders
of magnitude in mass, from 
$10^{12}M_{\sun}$ to $10^{15}M_{\sun}$.   
The resulting lensing signal is consistent with an NFW or Einasto
profile on scales outside the central region. In the inner regions,  
uncertainty in modeling of the proper identification of the 
halo center and inclusion of baryonic effects from the
central galaxy make the comparison less reliable. We find that the NFW concentration parameter $c_{200b}$
decreases with halo 
mass, from around 10 for galactic halos to 4 for cluster halos. 
Assuming its dependence on  halo mass in the form of 
$c_{200b} = c_0 (M/10^{14}\hmsun)^{-\beta}$ 
we find $c_0=4.6 \pm 0.7$ (at $z=0.22$) and $\beta=0.13\pm 0.07$, with 
very similar results for the Einasto profile. 
The slope ($\beta$) is in agreement with theoretical predictions, while the amplitude
is about two standard deviations
below the predictions for this mass and redshift, 
but we note that the published values in the literature differ at a level of 10-20\% and 
that for a proper comparison our analysis should be repeated in simulations. 
We compare our results to other recent determinations, some of which
find significantly higher concentrations. 
We discuss the  implications of our results for the baryonic
effects on the shear power spectrum: since these are expected to increase
the halo concentration, the fact that we see no evidence of high
concentrations  on scales above $20$\% of the virial 
radius suggests that baryonic effects are 
limited to small scales, and are not a significant source of uncertainty for
the current weak lensing measurements of the dark matter power spectrum. 

\end{abstract}

\maketitle

\section{Introduction}

The density profile of dark matter (DM) halos 
is one of the fundamental predictions of cosmological 
models in the nonlinear regime, determined 
using N-body simulations
\citep{1996ApJ...462..563N,1997ApJ...477L...9F,1997ApJS..111...73K,1998ApJ...499L...5M,1999MNRAS.310..527A,2000ApJ...544..616G,2000ApJ...529L..69J,2001MNRAS.321..559B,2001ApJ...554..903K,2001ApJ...557..533F,2002ApJ...568...52W,2003ApJ...588..674F,2003ApJ...597L...9Z,2004ApJ...607..125T,2005MNRAS.364..665D}.
The profile is often parametrized with the so called NFW profile
\citep{1996ApJ...462..563N}, a broken power-law 
characterized by a scale radius where the slope is approximately $-2$, and
by the virial radius which parametrizes its mass.
The ratio of the latter to the former is called the concentration $c$, and is a measure
of the density of the halo in the inner regions: a higher
concentration implies   
a higher density of the halo at a fixed fraction of virial radius. 
The concentration is predicted to be mildly dependent on mass such 
that higher mass halos are less concentrated than lower mass halos. 
Because the normalization of the concentration-mass relation depends on the matter power spectrum
normalization, its shape, and the matter density $\Omega_m$, measuring
the halo profile as a function of halo mass 
can teach us about the underlying cosmology.

Many different methods have been used to measure the halo profile of 
clusters.  Kinematic tracers such as satellite galaxies, in
combination with a Jeans analysis or  caustics analysis, can give
information over a wide range of physical scales and halo masses. 
While the issues of relaxation, velocity bias, anisotropy of 
the orbits and interlopers continue to be debated and need to be carefully addressed, 
recent results suggest a good agreement with theoretical predictions 
\citep{2003ApJ...585..205B,2004ApJ...600..657K,2003AJ....126.2152R,2005ApJ...628L..97D,2006AJ....132.1275R,2007MNRAS.378...41S}.  
Hydrostatic analyses of X-ray intensity profiles of clusters 
use X-ray intensity and temperature as a
function of radius to reconstruct the density profile.
They have the benefit of thermal gas pressure being isotropic, but 
may be biased due to the possible presence of other
sources of pressure support, such as 
turbulence, cosmic rays or magnetic fields. These cannot be strongly
constrained for typical clusters with present X-ray data \citep{2004A&A...426..387S}, but 
could modify the hydrostatic equilibrium and affect the conclusions of such
analyses. Recent results are encouraging and are in a broad agreement with 
predictions, although most
require concentrations 
that are higher than those predicted by a concordance cosmology
\citep{2006ApJ...640..691V,2007ApJ...664..123B,2007MNRAS.379..209S}. 
While the above-mentioned systematic biases cannot be excluded, the small discrepancy could also 
be due to baryonic effects in the central regions, 
due to selection 
of relaxed clusters that may be more concentrated than average \citep{2006ApJ...640..691V}, or due
to the fact that at a given X-ray
flux limit, the more concentrated clusters  near the
limiting mass are more likely to be included in the sample
\citep{2007A&A...473..715F}.

Gravitational lensing is by definition sensitive to the total mass, and is
therefore one of the most promising methods to measure the mass profile. 
Some analyses have combined strong- and weak-lensing or velocity dispersion 
constraints for individual clusters \citep{2005ApJ...621...53B,
2006ApJ...640..639Z,2007ApJ...668..643L,2006ApJ...652..937B,2004ApJ...604...88S,2007MNRAS.379..190C}
to derive the profile, and
in some cases concentrations have been found above the predictions from simulations. 
However, strong lensing is affected by the mass distribution 
in the very inner parts of the cluster, and both ellipticity of matter and 
stars from the central galaxy 
have a significant effect on the strong lensing signatures, so that 
these analyses are not necessarily measuring the 
primordial dark matter halo profile \citep{2007MNRAS.381..171M}.
Furthermore, both strong lensing and X-ray analyses are susceptible to
selection bias effects when changing the DM concentration at fixed
mass \citep{2007A&A...473..715F}.    
The problems in interpreting the observations on small scales, where baryons 
play an important role, suggest that we should focus on larger scales if we 
want to compare observations to theoretical predictions from N-body simulations. 
Weak lensing is arguably the most promising tool that can be used to measure the profile 
out to scales of several \hmpc. It has the advantage that 
outside the central region, the dark matter distribution is likely to be unaffected 
by baryons, so a comparison against N-body simulations should be more reliable.  

Many previous weak lensing analyses have focused on individual clusters
(for example, \citep{2007MNRAS.379..317H,2007ApJ...667...26P}).
Measuring the matter distribution of individual clusters has its advantages, since 
it allows a comparison with the light and gas distributions on an 
individual basis, and so can  constrain models that relate the two, such as MOND versus CDM 
\citep{2006ApJ...648L.109C}. 
However, since lensing measures a projected 
surface density with a window that extends hundreds of megaparsecs away, 
other mass perturbation along the 
line of sight will also produce a lensing signal and thus 
act as a source of noise when extracting
the cluster density profile.
Some of these structures may be 
correlated with the cluster itself, for example those that are 
falling into the cluster along the
filaments connected to the cluster \citep{2001ApJ...547..560M}, and
can be defined as part of the cluster   
profile, while other structures may be 
completely unrelated mass concentrations tens or hundreds of megaparsecs away \citep{2003MNRAS.339.1155H}. 

The measurement of the dark matter profile can therefore be quite noisy for individual
clusters.  Stacking the signal from many clusters can ameliorate this
problem, since
only the mass density correlated with the cluster will produce a
signal. 
This way, the measurement determines the true average cluster mass
profile in the same way as defined
in simulations. Such a statistical approach is thus advantageous if
one is to compare the
observations to theoretical predictions, which also average over a large
number of halos in simulations. In fact, simulations show a
significant scatter in the shapes of individual halo profiles
\cite{2001MNRAS.321..559B,2007MNRAS.381.1450N},  so
stacking  many halos will reduce the fluctuations due to noise
caused by uncorrelated
structures along the line of sight, due to shape measurement noise,
and due to the shape variations of 
individual halos.  A final advantage of stacking is that it allows
for the lensing measurement of
lower-mass halos, where individual detection is impossible due to
their lower shears relative to
clusters.  Individual high signal-to-noise cluster observations and those based on
stacked analysis of many clusters are thus complementary to each other
at the high mass end, with the stacked analysis drastically increasing
the available baseline in mass.

The statistical approach based on stacked clusters
has been applied to a small number of clusters before
\cite{2003ApJ...588L..73D,2001ApJ...554..881S}. The Sloan Digital Sky Survey (SDSS)
provides an ideal dataset for such analysis:
it covers a
significant fraction of the sky containing $\sim 10^4$ clusters up to
$z\sim 0.3$, providing a large
volume of the universe over which the clusters can be observed.
The SDSS spectroscopy and multicolor imaging enables
precise redshift determination, so we can
determine the profile as a function of true transverse separation rather than
angle.
In a previous analysis, we have used a sample of 43~335 groups and clusters
as traced by luminous red galaxies (LRGs) \citep{2006MNRAS.372..758M} to derive
average mass density profiles of groups with masses between
$3\times 10^{13}\hmsun$ and $1.3\times
10^{14}\hmsun$. We have also performed a halo model analysis of
the lensing signal of isolated $\sim L_*$ elliptical galaxies in 
\cite{2006MNRAS.368..715M} to ensure that they are all in the field,
and will augment that sample with isolated $\sim L_*$ spiral
galaxies.  In this paper
we extend these previous analyses to the new sample of 13~823 MaxBCG clusters
presented in \cite{2007ApJ...660..239K}, which extends the mass range
to $\sim 6\times 10^{14}$ \hmsun. We also compare our results against
an independent analysis of these clusters in 
\cite{2007arXiv0709.1159J}, though both analyses have
included objects other than the public maxBCG catalog selected in
different ways. We then combine the elliptical,
spiral, LRG and MaxBCG analyses to obtain
information about the halo density profile over a wide range of masses,
and compare them to theoretical predictions from N-body simulations.

We begin in section~\ref{S:theory} by presenting the theory behind our
measurement, the data, and the analysis method used.  The results are
presented in section~\ref{S:results}, including a discussion of how to
compare them with other observations and with theory.  Our conclusions
derived from this analysis and comparison with theory are given in
section~\ref{S:conclusions}.  

\section{Data and analysis}\label{S:theory}

\subsection{Theory}\label{SS:theory}

We follow the same methodology as in \cite{2006MNRAS.372..758M}, so we
refer the reader
to that paper for a more detailed description of the analysis process.
In brief, cluster-galaxy and galaxy-galaxy weak lensing provide a
simple way to probe the
connection between clusters (or galaxies) and matter via their
cross-correlation functions $\xi_{cl,m}(\vec{r})$ (or $\xi_{g,m}(\vec{r})$),
which can be related to the
projected surface density
\beq\label{E:sigmar}
\Sigma(R) = \overline{\rho} \int \left[1+\xi_{cl,m}\left(\sqrt{R^2 + \chi^2}\right)\right] d\chi
\eeq
where $R$ is the transverse separation and $\chi$ the radial direction
over which we are projecting.  We are ignoring the effects from the 
radial window, which is hundreds of megaparsecs broad
and not relevant at cluster scales. The surface density is then related to the observable
quantity for lensing, the differential surface density, 
\beq\label{E:ds}
\ds(R) = \gamma_t(R) \Sigma_c= \overline{\Sigma}(<R) - \Sigma(R), 
\eeq
where the second relation is true only in the weak lensing limit, for
a matter distribution that 
is axisymmetric along the line of sight (which is naturally achieved
by our procedure of stacking thousands of clusters to determine their
average lensing signal).  This observable quantity can
be expressed as the product of two factors, a tangential shear
$\gamma_t$ and a geometric factor
\beq\label{E:sigmacrit}
\Sigma_c = \frac{c^2}{4\pi G} \frac{D_S}{D_L D_{LS}(1+z_L)^2}
\eeq
where $D_L$ and $D_S$ are angular diameter distances to the lens and
source, and $D_{LS}$ is the angular diameter distance between the lens
and source. 
Unless otherwise noted, 
all computations assume a flat $\Lambda$CDM universe with
$\Omega_m=0.27$ and $\Omega_{\Lambda}=0.73$.  Distances
quoted for 
transverse lens-source separation are comoving (rather than physical)
\hkpc, where $H_0=100\,h$ km$\mathrm{s}^{-1}\,\mathrm{Mpc}^{-1}$.
Likewise, the differential surface density 
\ds{} is computed in
comoving coordinates, and 
the factor of $(1+z_L)^{-2}$ arises due to our use of
comoving coordinates.

For this paper, we are primarily interested in the contribution to the
cluster-mass or galaxy-mass cross-correlation from the halo profile itself,
rather than from neighboring halos (halo-halo term), and hence 
\beq\label{E:sigmar2}
\Sigma(R) = \int_{-\infty}^{\infty} \rho(r=\sqrt{\chi^2+R^2}) d\chi
\eeq
The halo-halo term for clusters in host halos can be modeled simply
using the cluster- or galaxy-dark matter cross-power spectrum as in, e.g.,
\cite{2005MNRAS.362.1451M}, and is only important for $R >\sim 2$
\hmpc. 
Nevertheless,  
we compute this component and include it in the model as a fixed term
which we obtain by computing first the mass of the clusters,
deriving the corresponding halo bias using the bias-mass relation
\citep{1999MNRAS.308..119S,2004MNRAS.355..129S} 
and using the linear power spectrum multiplied with the halo bias 
to obtain the halo-halo term. This procedure has been shown to work 
well in comparison to simulations \citep{2005MNRAS.362.1451M}, so we
use the bias-mass relation from that paper but with 
$\sigma_8=0.75$ and $\Omega_m=0.27$, computed at  the mean redshift of each
sample.  

We can model the one-halo term for each sample 
as a sum of the stellar component, only important 
on scales below $\sim 100$\hkpc, and an NFW 
dark matter profile \citep{1996ApJ...462..563N}:
\begin{equation}\label{E:genprofile}
\rho(r) = \frac{\rho_s}{\left(r/r_s\right)
  \left(1+r/r_s\right)^{2}}. 
\end{equation}
  It is convenient to reparametrize it by two  
  parameters, concentration 
$c=r_{vir}/r_s$ and virial mass $M$.  The virial radius $r_{vir}$
  and $\rho_s$ can be 
  related to $M$ via consistency relations.  The first is that the
  virial radius is that within which the average density is equal to
  $200$ times the mean density:
\begin{equation}\label{E:rvir}
M_{200b} = \frac{4\pi}{3}r_{vir}^3 \left(200\overline{\rho}\right).
\end{equation}
The second relation, 
used to determine $\rho_s$ from $M$ and $c$, is simply that the
volume integral of the density profile to the virial radius must
equal the virial mass.
The NFW 
concentration $c$ is a weakly decreasing function of halo mass, with a 
typical dependence as 
\begin{equation}\label{E:cmrelation}
c_{200b}=\frac{c_0}{1+z} \left( \frac{M}{M_0} \right)^{-\beta}, 
\end{equation}
with $\beta \sim 0.1$
\citep{2001MNRAS.321..559B,2001ApJ...554..114E,2007MNRAS.381.1450N}, making this
profile a one-parameter family of 
profiles. The normalization depends on the nonlinear mass, which is
the mass within spheres  
in which the rms fluctuation in the linear regime is 1.68; 
for the typical range of cosmological models, one expects $c\approx 5-6$ at
$M_0=10^{14}\hmsun$.  

\begin{figure}
\begin{center}
\includegraphics[width=3in,angle=0]{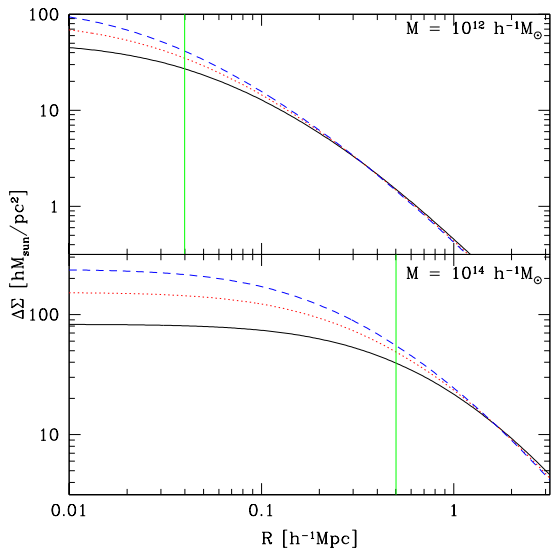}
\caption{\label{F:diffc}The NFW profile theoretical lensing signal \ds\ 
  for two mass scales.  In the upper plot, with
  $M=10^{12}h^{-1}M_{\odot}$, the concentration is varied from 7 to 10
  to 13.  In the lower panel, with $M=10^{14}h^{-1}M_{\odot}$, it is
  varied from 3 to 5 to 7. The curves on each panel are normalized to
  the same virial mass $M_{200b}$.  Vertical lines show the minimum
  scale for our fits.}
\end{center}
\end{figure}

As an example of how the lensing signal varies with concentration, in
Fig.~\ref{F:diffc} we
show the predicted NFW lensing signal as a function of transverse separation for
$M_{200b}=10^{12}$ (galaxy scale) and $10^{14} h^{-1}M_{\odot}$ (cluster scale) halos, for several plausible
concentration values.  The vertical line shows the minimum scale used
for our fits, and it is clear that the lensing signal above those scales
can differentiate between different concentration values.  
This discriminating power stems in part from the fact that the lensing signal
reflects the differential surface density, which draws information
from smaller scales to larger scales.

We fit the data to the model assuming a spherical NFW profile. 
In \cite{2005MNRAS.362.1451M}, it was shown
that spherical NFW profiles do an excellent job at describing the stacked 
lensing signal from simulations for a variety of masses, with the masses and
concentrations of the best-fit profiles related to the real masses and
concentrations in the simulations in a particular way (to be discussed
further below).  The fit $\chi^2$ values were very good when the
errorbars used on the simulated lensing signal were one-tenth of our current errorbars,
so henceforth we consider only spherical NFW fits rather than trying
to account for the averaging of triaxial halos.

Although we could include the effects of the central galaxy to make the model 
accurate on scales below 20\% of the virial radius, there are reasons why we should 
exclude this information from the fits if we want to compare against N-body simulations. 
First, the baryonic effect we assume may not 
be completely accurate because of uncertainties in the stellar mass to 
light ratio, and in the dark matter response to the presence of baryons and stars forming out of them. 
The latter 
is often modeled as adiabatic contraction
\cite{1986ApJ...301...27B,2004ApJ...616...16G}, but this prescription
may inaccurately describe 
the actual effect, depending on the formation history of the
galaxy, group, or cluster 
in question \citep{2007ApJ...658..710N}. The second, possibly more
important effect is that  
for cluster samples such as MaxBCG, the cluster center cannot always be reliably
determined using the optical information.  This uncertainty has two
causes: first, that a brightest cluster galaxy (BCG) may not be located at
the deepest part of the 
cluster potential, but may be offset from it due to (for example) perturbations
from infalling satellites; and second (and more importantly), that the
maxBCG algorithm may choose the wrong BCG.   Studies comparing the BCG
position to the center defined by either X-ray intensity or by average  
satellite velocity have found that the typical 
displacement is about 2-3\% of the virial radius when the BCG is properly
identified \citep{2005MNRAS.361.1203V,2007ApJ...660..221K,2007ApJ...660..239K,2008arXiv0802.2712B}. 
The last of these studies finds that for about 10\% of BCGs, the displacement extends to above 10\% of the virial radius. 
It is interesting to note that selecting blue core BCGs significantly reduces the displacement to below 1\%
of the virial radius \citep{2008arXiv0802.2712B}. However, while the
multi-band SDSS aperture photometry could be used to select blue-core
BCGs, we do not undertake this approach for this analysis because of
decreased $S/N$.  

Another study that includes red galaxy photometric errors  (i.e., both
effects rather than just the first) finds that
the median displacement is 10\% of the virial radius \citep{2007arXiv0706.0727H}.
 If the 
assumed center is displaced from the true center, then this can have a 
significant effect on the density distribution in the inner parts, mimicking 
a halo profile with a lower concentration
\citep{2002MNRAS.335..311G,2006MNRAS.373.1159Y}, while in the outer
parts the effect  is significantly smaller.  To avoid this problem, we
use a minimum scale for the fits of 500 \hkpc\ for the maxBCG clusters
(typically half the
virial radius), and present evidence that this approach is robust to
the effects of centroiding problems, at the expense of decreased
statistical power. 

 We note that this problem has also been addressed
in a statistical manner,  
using mock catalogs to determine both the fraction of clusters affected by
centroiding problems, and also the distribution of projected offsets from
the true cluster center for those that are centroided wrong
\citep{2007arXiv0709.1159J}.  Using this correction, one can in
principle correct for the effect and fit to significantly smaller
scales; however the correction is quite dependent on the content of
the mock catalogs, and any deviation from reality may invalidate it.
For example, the mock catalogs suggest that misidentification of the center is 
more of a problem at lower halo masses, while our 
visual inspection of clusters suggests the opposite, in the sense that 
the most massive clusters (with richness $N_{200}>80$) have 
significant bimodality and substructure, as expected from a 
hierarchical cosmological model where the most massive clusters formed the latest. 
As a result, we exclude these most massive clusters
from the analysis entirely. For the remaining sample,
we choose the different approach of avoiding the inner parts of the cluster 
(where the effect is strongest), thereby decreasing the systematic error at the 
expense of an increase in statistical error.  

\subsection{Data}

The data used here are obtained from the SDSS
\citep{2000AJ....120.1579Y}, an ongoing survey to image roughly
$\pi$ steradians of the sky, and follow up approximately one million of
the detected objects spectroscopically \citep{2001AJ....122.2267E,
2002AJ....123.2945R,2002AJ....124.1810S}. The imaging is carried out
by drift-scanning the sky 
in photometric conditions \citep{2001AJ....122.2129H,
2004AN....325..583I}, in five bands ($ugriz$) \citep{1996AJ....111.1748F,
2002AJ....123.2121S} using a specially-designed wide-field camera
\citep{1998AJ....116.3040G}. These imaging data are used to create the
source catalog that we use in this paper. In addition, objects are targeted for
spectroscopy using these data \citep{2003AJ....125.2276B} and are observed
with a 640-fiber spectrograph on the same telescope
\citep{2006AJ....131.2332G}. All of the data are 
processed by completely automated pipelines that detect and measure
photometric properties of objects, and astrometrically calibrate the data
\citep{2001ASPC..238..269L,
  2003AJ....125.1559P,2006AN....327..821T}. The SDSS has had seven
major data releases \citep{2002AJ....123..485S,
2003AJ....126.2081A, 2004AJ....128..502A, 2005AJ....129.1755A,
2004AJ....128.2577F, 2006ApJS..162...38A,2007ApJS..172..634A,2008ApJS..175..297A}. 

In the subsections that follow, we describe the lens and source samples.

\subsubsection{MaxBCG cluster lenses}

Our highest mass lens sample consists of 13~823 MaxBCG clusters 
\citep{2007ApJ...660..221K,2007ApJ...660..239K}, which are identified 
by the concentration of galaxies in color-position space, using 
the well known red galaxy color-redshift
relation \citep{2000AJ....120.2148G}. The sample is based on 7500 square degree of imaging data in SDSS. 
There is a tight mass-richness relation that 
has been established using dynamical information 
across a broad range of 
halo mass \citep{2007ApJ...669..905B}. This is also confirmed with the
lensing analysis, as shown below and in
\citep{2007arXiv0709.1153S,2007arXiv0709.1159J,2008arXiv0802.2365R}.   

The redshift range of the MaxBCG catalog is 
$0.1<z<0.3$, and the upper cutoff ensures that the
lenses still have a sufficient number of sources behind them.  Within
these redshift limits, the sample is approximately volume-limited with 
a number density of $3\times 10^{-5} (h/\mbox{Mpc})^3$, except for a
tendency towards higher number density at the lower end of this 
redshift range \citep{2008arXiv0802.2365R}. 
The main tracers of halo mass provided by the MaxBCG team are a
rescaled 
richness $N_{200}$ (number of red galaxies above $0.4L_*$), total
luminosity $L_{200}$ (the luminosity in those red galaxies, including
the BCG) and 
BCG luminosity $L_{\rm BCG}$. 

In this paper we use richness $N_{200}$ as a primary tracer 
of halo mass.  However, we observe (based on \cite{2008arXiv0802.2365R}) that the lower mass end of the
public catalog overlaps with the LRG samples used for this work, so we
eliminate those clusters with the lowest richness, $N_{200}=10$ and $11$ (1/3 of the public catalog),
and split the remainder into six narrow $N_{200}$ bins as shown in
Table~\ref{T:lenses} (except for those with $N_{200}>80$, as described
in Sec.~\ref{SS:theory}).  The 
widest of these spans a factor of 1.4 in $N_{200}$, so if we assume
the mass is proportional to this observable, then the mass bin is a
factor of $1.4$ wide without any mass-observable scatter.  In reality
such scatter exists, but even a factor of two scatter in the mass at
fixed $N_{200}$ would give a mass distribution less than an order of
magnitude wide.  As we have shown in \citep{2005MNRAS.362.1451M}, when
fitting for a single halo
mass on a stacked sample that is less than an order of magnitude wide,
the best-fit mass is a good proxy (within $\sim 10$\%) for the true
mean mass of the sample.

\begin{table*}
\caption{\label{T:lenses}Summary of all lens samples used in this
  paper.  Note that observable definitions (such as luminosity) are not necessarily the
  same for each parent sample; see the relevant subsection for
  details.  Additional sample cuts are indicated by the parent
  samples, which typically are defined using flux and color cuts.}
\begin{tabular}{lcccccc}
Parent sample & Isolation method & Observable cut & Redshift cut &
$\langle z\rangle$ & $N$ & Name\\
\hline
\hline
MaxBCG clusters & MaxBCG method & $12\le N_{200}\le 13$ & $0.1<z<0.3$
& $0.22$ & $2531$ & \\
 &  & $14\le N_{200}\le 19$  &  & & $3372$ & \\
 &  & $20\le N_{200}\le 28$  &  & & $1618$ & \\
 &  & $29\le N_{200}\le 39$  &  & & $614$ & \\
 &  & $40\le N_{200}\le 54$  &  & & $248$ & \\
 &  & $55\le N_{200}\le 79$  &  & & $109$ & \\
\hline
Spectroscopic LRGs & Cylindrical & $M_{^{0.0}r} > -22.3$ &
$0.15<z<0.35$ & $0.24$ & $27700$ & \\
 &  & $-22.3 \ge M_{^{0.0}r} > -22.6$ & &  & $10536$  & \\
\hline
MAIN spectroscopic & Cylindrical & $-20\le M_{^{0.1}r} < -19$,
frac\_deV$\ge 0.5$ & $z>0.02$ & $0.07$ & $20150$ & L3 elliptical\\
 &  & $-21\le M_{^{0.1}r} < -20$, frac\_deV$\ge0.5$ &  & $0.10$ &
 $46130$ & L4 elliptical\\
 &  & $-21.5\le M_{^{0.1}r} < -21$, frac\_deV$\ge0.5$ &  & $0.13$ &
 $23485$ & L5faint elliptical \\
 &  & $-20\le M_{^{0.1}r} < -19$, frac\_deV$<0.5$ &  & $0.07$ &
 $38640$ & L3 spiral\\
 &  & $-21\le M_{^{0.1}r} < -20$, frac\_deV$<0.5$ &  & $0.10$ &
 $42235$ & L4 spiral \\
\hline
\hline
\end{tabular}
\end{table*}

\subsubsection{Spectroscopic LRGs}

For the next sample with lower average halo mass, we use 
the spectroscopic 
Luminous Red Galaxy (LRG) sample
\citep{2001AJ....122.2267E}, including area beyond Data Release 4
  (DR4). 
The total area coverage for this spectroscopic sample is 5154 square
  degrees, as 
available in the NYU Value Added Galaxy
Catalog (VAGC, \cite{2005AJ....129.2562B}) at the time of the
  original publication of our lensing work with this sample
  \citep{2006MNRAS.372..758M}.   

For the LRG sample, we define luminosities using $r$-band model
magnitudes, extinction-corrected using reddening maps from
\cite{1998ApJ...500..525S}.  We 
apply a k+e-correction (combined $k$-correction and correction for
evolution of the spectrum) to redshift zero 
using stellar population synthesis code from
\cite{2003MNRAS.344.1000B}; this magnitude is denoted $M_{^{0.0}r}$ to
distinguish it from the magnitudes used for the Main sample, which are
corrected to $z=0.1$.  In our original work, the LRGs were split
into two luminosity bins, $M_{^{0.0}r} \ge -22.3$ and 
$M_{^{0.0}r} < -22.3$.   To reduce the overlap
between the LRGs and the cluster samples to $<5$\%, we use the fainter
bin in its entirety but only use the $-22.6\le M_{^{0.0}r} < -22.3$
subset of the brighter sample.  

These LRG samples were derived from
the full LRG sample after eliminating 
$15$\% of the sample using a cylindrical density estimator, designed to
avoid satellite galaxies due to their extra lensing signal from the
host halo.  Specifically, the LRGs were each required to be the only
or the brightest LRG in a cylinder with radius $R=2h^{-1}$Mpc and
line-of-sight separation $\delta v=\pm 1200$ km s$^{-1}$.  This cut is
conservative, in the sense that for typical groups and low-mass
clusters, we may have excluded some host galaxies; however, the host
sample purity is sufficiently important for this analysis that we tend
towards the conservative side.  More
information about the LRG samples is available in Table~\ref{T:lenses}. 

\subsubsection{$L_*$ lenses}

Finally, we include
lower luminosity samples from \cite{2006MNRAS.368..715M} that have
been shown with a robust environment estimator to consist of
field galaxies. For those samples, we used galactic extinction-corrected
$r$-band Petrosian magnitudes, $k$-corrected to $z=0.1$ using {\sc
  kcorrect v4\_1\_4} \citep{2003AJ....125.2348B}, denoted $M_{^{0.1}r}$.

 The samples used here correspond to L3, L4, and
  L5faint 
isolated ellipticals from \cite{2006MNRAS.368..715M}, where the isolated
ellipticals are the half of the elliptical sample at those
luminosities determined to be in the field using a cylindrical density
  estimator. We also include a sample not shown
there but from the same data, of L3 and L4 isolated spirals, or 85\%
of the spirals in those luminosity bins.  For reference, $L_*$ is
  within L4, so our samples range from slightly below to slightly
  above $L_*$.  In this context, the
elliptical and spiral samples are chosen using the SDSS frac\_deV
parameter, which determines whether the light profile is closer to a
de Vaucouleurs or exponential profile.  More details of these five
  lens samples are shown in Table~\ref{T:lenses}.

We do not use the fainter samples (L1 and L2) from previous work
because the detection significance of the weak lensing signal is
low, so they cannot constrain the halo concentration.  We also
avoid the brighter elliptical samples because they overlap with the spectroscopic
LRGs, and the brighter spiral samples because they are nearly empty.

\subsubsection{Lensing sources}

The source sample used is the same as that originally described in
\cite{2005MNRAS.361.1287M}.  This source sample
includes over 30 million galaxies from the SDSS imaging data with
$r$-band model magnitude brighter than 21.8, with
shape measurements obtained using the REGLENS pipeline, including PSF
correction done via re-Gaussianization \citep{2003MNRAS.343..459H} and
with cuts designed to avoid various shear calibration biases. 
In addition, there are also uncertainties due to 
photometric redshifts and/or redshift distributions
of background galaxies, which were originally  calibrated using DEEP2 Groth strip data, 
as well as due to other issues affecting the
calibration of the lensing signal, such as the sky subtraction uncertainties,
intrinsic alignments, magnification bias, star-galaxy separation, and
seeing-dependent systematics. The 
overall $1\sigma$ calibration uncertainty was estimated to be eight per cent
\cite{2005MNRAS.361.1287M}, though the redshift calibration component
of this systematic error budget has recently been significantly decreased due to
the availability of more spectroscopic data \citep{2008MNRAS.386..781M}. 
The calibration mainly affects the mass estimation rather than the 
derived density profiles, so it is not of significant concern for 
this paper due to the weak dependence of concentration on mass. 

An additional concern is the relative calibration of lensing
measurements from different lens samples that were published in
different papers.  If there are calibration differences
between these measurements, then the power-law scaling of the $c(M)$
relation might be misestimated.  However, it seems unlikely that there can be
significant calibration differences between the different
measurements, for two reasons.  First, the same version of
the source catalog was used for each one.  This suggests that any
shear calibration or star/galaxy separation issues are the same for
each measurement.  There may be very slight variation due to the
different mean lens redshifts, which changes the effective mean redshift
of the sources, but our previous tests of the source catalog for
relative shear calibration as a function of apparent magnitude and
size \citep{2005MNRAS.361.1287M} rule out changes in the calibration
of the shear that are 
significant relative to the $1\sigma$ statistical error on these
measurements.  Second, we have rigorously tested the calibration of
the source redshift distributions using the zCOSMOS and DEEP2
spectroscopic samples, and found that the calibrations for the
lens redshift distributions used here are the same within several
percent, which is again smaller than the $1\sigma$ measurement error on the
lensing signals used here \citep{2008MNRAS.386..781M}.  In short,
while there is some small uncertainty (discussed above) in the
absolute lensing signal calibration, we have little reason to believe there is
any significant discrepancy between the calibrations for the different
lens subsamples.

\subsection{Analysis}\label{SS:analysis}

Calculation of the signal is described in detail in \cite{2006MNRAS.372..758M}. Briefly, we 
compute the weights based on noise and redshift information for each 
lens-source pair, summing them using a minimal variance estimator. 
We compute the signal
around random points and subtract it from the signal
around real lenses to eliminate contributions from systematic shear.
The signal must be boosted, i.e. multiplied by $B(R) =
n(R)/n_{rand}(R)$, the ratio of the number density of sources relative
to the number around random points, in order to account for dilution
by sources that are physically associated with lenses, and therefore
not lensed. The former correction is only important on scales above 
those used in this paper ($>5$ \hmpc) and the latter on scales below
$20$\% of the virial radius, which we do not use for the fits in this paper. 
To determine errors on the lensing signal, we divide the
survey area into 200 bootstrap subregions, and generate 2500
bootstrap-resampled datasets.  
We note that the effects of non-weak shear, magnification bias, 
sky subtraction and 
intrinsic alignments, discussed in more detail in
\cite{2005MNRAS.361.1287M,2006MNRAS.372..758M}, are 
negligible on the scales used in this paper. 

The errors determined from the bootstrap are used in plots of the
signal with errors; however, the bootstrap covariance matrices can be
quite noisy and therefore inappropriate to use for weighting in the
$\chi^2$ minimization fitting.  To avoid this problem, we determine
analytic, diagonal covariance matrices (including shape noise), which
we have 
shown in \cite{2005MNRAS.361.1287M} to be a less noisy version of the bootstrap
covariance matrices with agreement in size of the errors at the $\sim
10$\% level.  These covariance matrices, which are far less noisy, are
used to perform the fits on each bootstrap-resampled dataset.  The
distributions of output parameters from all the bootstrap-resampled
datasets are used to determine errors on the fit parameters.

The fits are for eleven parameters, using the lensing signal from the
thirteen lens samples in Table~\ref{T:lenses}:
\begin{itemize}
\item The normalization and slope of the concentration-mass relation,
  Eq.~\ref{E:cmrelation} (2 parameters).
\item The normalization and slope of the relation between mass
  $M_{200b}$ and MaxBCG richness $N_{200}$ (2 parameters):
\beq
M_{200b} = M_0 \left( \frac{N_{200}}{20} \right)^{\gamma}
\eeq
\item The masses of the two LRG and five lower luminosity samples (7 parameters).  
\end{itemize}

As shown in Eq.~\ref{E:cmrelation}, the concentration is expected to
scale with redshift.  Since the samples are at different mean
redshifts, we use the expected redshift scaling to fit for a
normalization $c_0(z=0.22)$ (the mean, lensing-weighted redshift of the maxBCG sample).

\section{Results: Lensing signal and fits to NFW profile}\label{S:results}

\begin{figure*}
\begin{center}
\includegraphics[width=6in,angle=0]{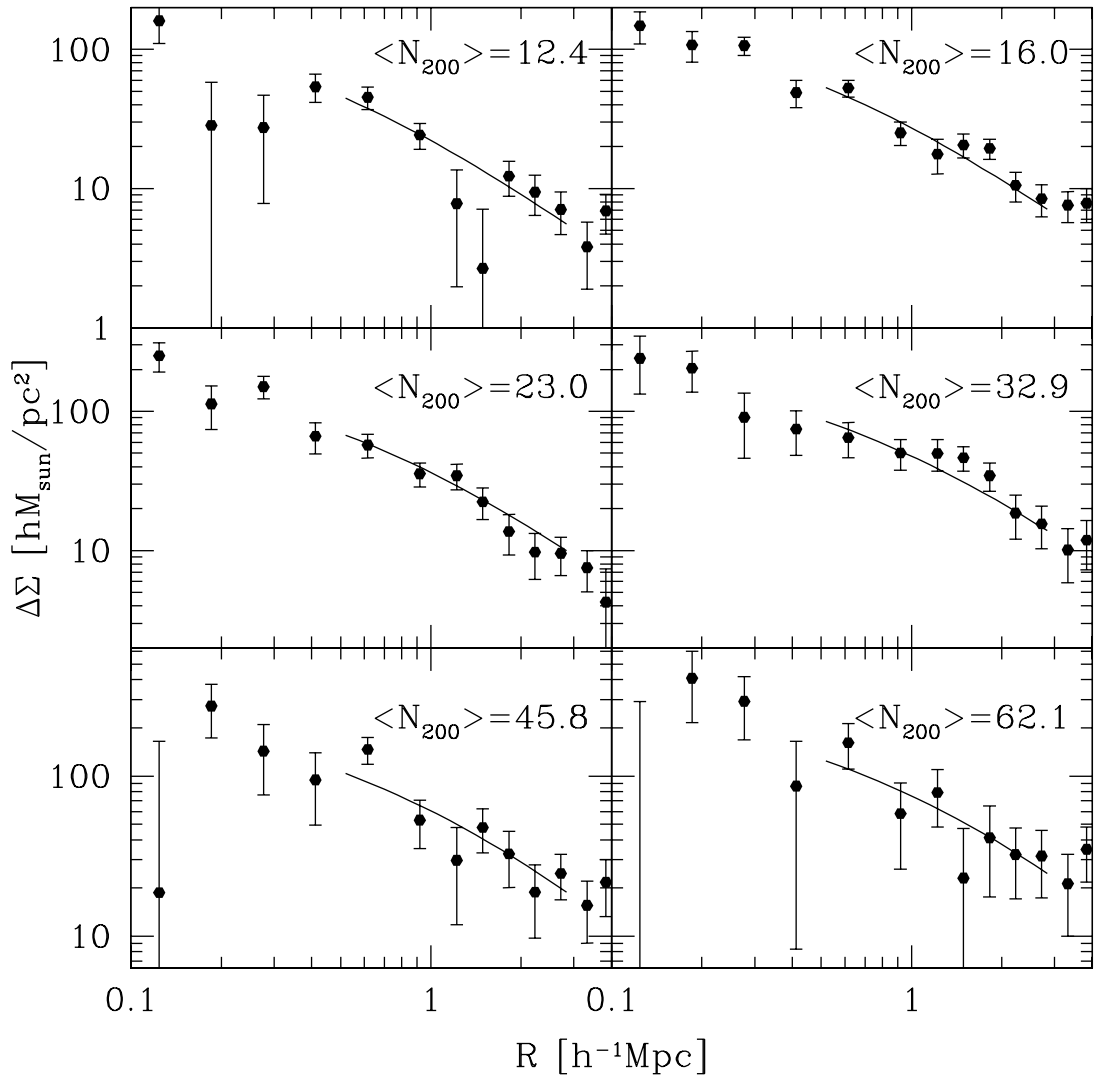}
\caption{\label{F:signal}The lensing signal \ds\ and model prediction
  for the best-fit model parameters (fit 2 in Table~\ref{T:cfits}), for the maxBCG sample split into
  6 richness bins.}
\end{center}
\end{figure*}
\begin{figure*}
\begin{center}
\includegraphics[width=6in,angle=0]{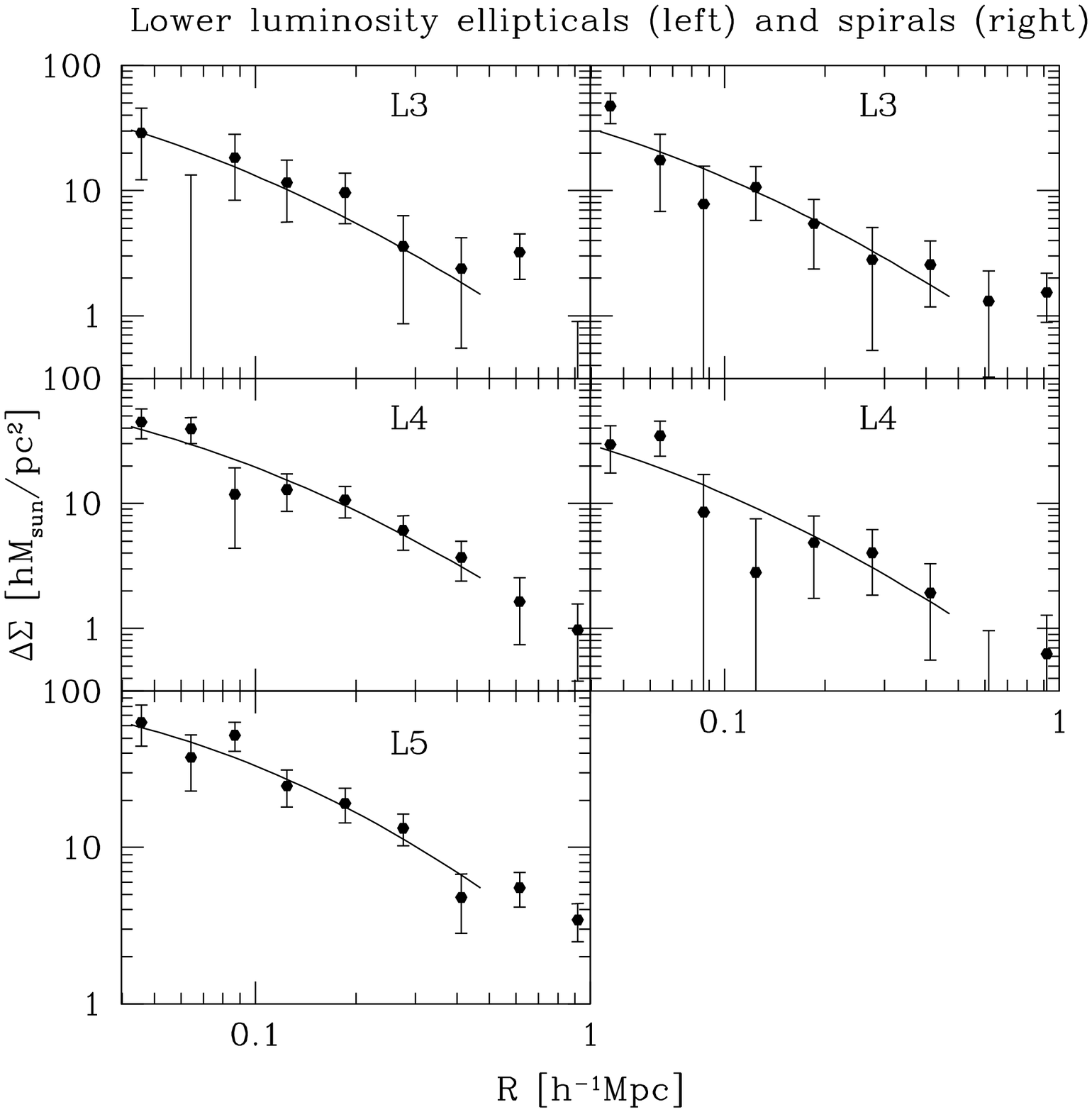}
\caption{\label{F:signallowlum}The lensing signal \ds\ and model prediction
  for the best-fit model parameters (fit 2 in Table~\ref{T:cfits}), for the field low luminosity sample split into
  5 morphology and luminosity bins (left column: ellipticals; right
  column: spirals).}
\end{center}
\end{figure*}
\begin{figure}
\begin{center}
\includegraphics[width=3.2in,angle=0]{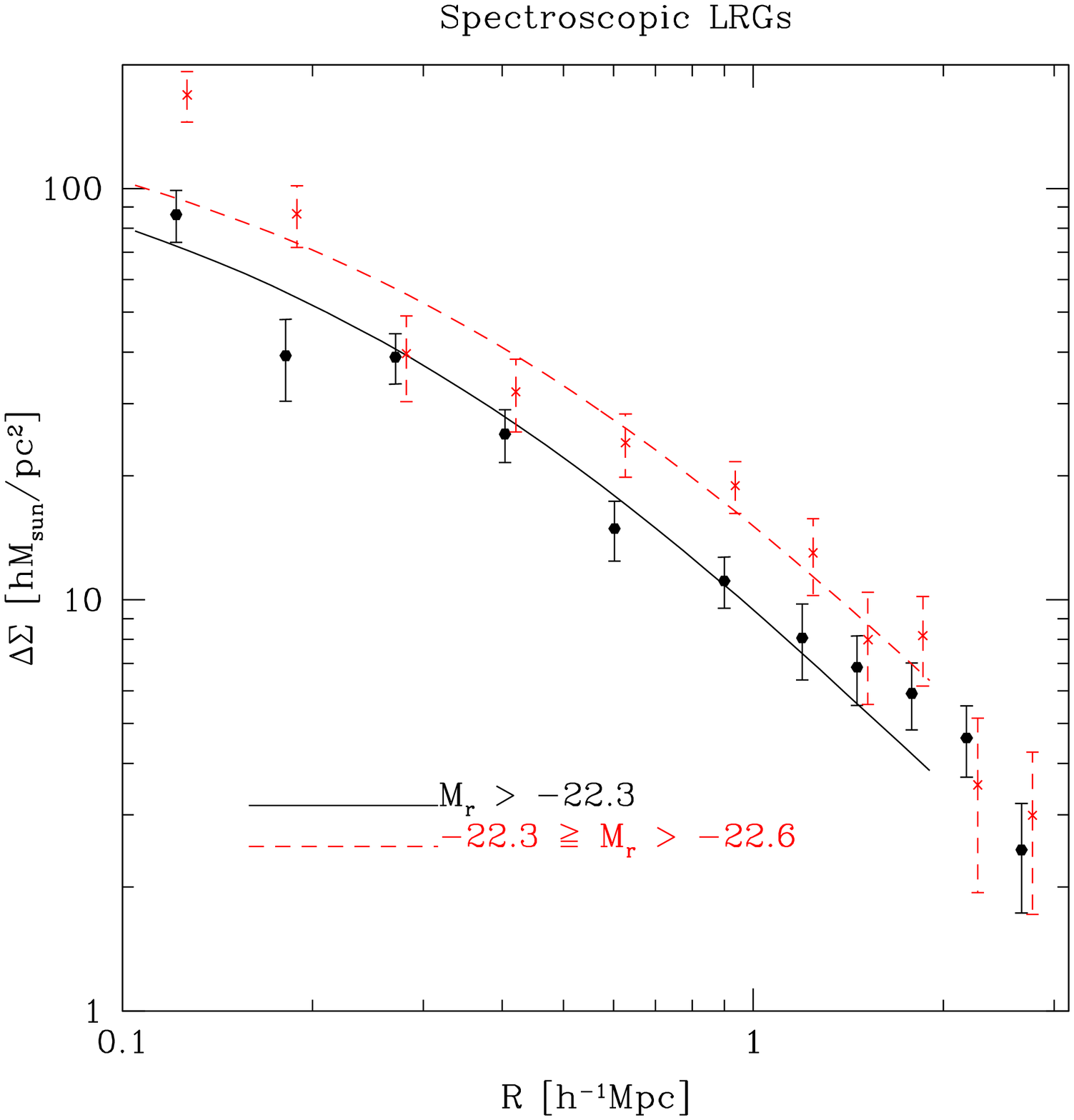}
\caption{\label{F:signalLRG}The lensing signal \ds\ and model prediction
  for the best-fit model parameters, for the host LRG sample in two
  luminosity bins as labeled on the plot.}
\end{center}
\end{figure}

We perform the analysis described in section~\ref{SS:analysis} on the lensing signal
\ds\ for all the lens samples.  Fig.~\ref{F:signal} 
shows the lensing signal \ds\ with the best-fit model in the joint
fits for $c_{200b}(M_{200b}, z=0.22)$, for
$M_{200b}(N_{200})$ for the maxBCG clusters, and best-fit masses for the
spectroscopic LRGs and the lower luminosity galaxy samples.
Fig.~\ref{F:signallowlum} shows the signal and best-fit model for the
lower luminosity samples, and
Fig.~\ref{F:signalLRG} shows the same for the two LRG samples.   
We see that for the maxBCG sample there is a strong lensing signal over the entire range 
of richness, and the lensing signal is increasing with richness as expected. 
The best-fit parameters for concentration and mass (to be described
below) clearly provide a good fit to the data in all lens samples, as
will also be evident in the $\chi^2$ values for the fits, discussed below.

\subsection{Description of fits}

Table~\ref{T:cfits} shows the best-fit parameters for the power-law
relations $c(M)$ and maxBCG $M(N_{200})$, using 
several approaches to the fits as described below.  The table includes
information about the fit minimum  radii and whether
offsets of BCGs were accounted for in the fitting. 
It also shows the $\chi^2$ per degree of freedom, indicating that all
fits shown provide a reasonable fit to the data.
As will be discussed in detail later, Fit 2 in that table is used for
any plots that include a best-fit model, because the fit procedure
represents the best possible trade-off between systematic error due to cluster
centroiding errors and statistical error.  Table~\ref{T:massfits}
shows the best-fit masses for the two LRG lens samples and the five
lower luminosity samples for fit 2 in Table~\ref{T:cfits}.

The tables include the results for several types of fits with varying
minimum radii for the maxBCG sample; these results were shown to test
several possible issues.  In all cases, the maximum fit radius for the
maxBCG clusters was 3.0 \hmpc; the fits for the LRGs used $0.1$--$2$
\hmpc; and the fits for the lower luminosity samples used
$0.04$--$0.5$ \hmpc.  
The maximum fit radius was chosen so that the halo-halo term remains small
compared to the one halo term. When computing the halo-halo term we 
assume the cosmological model and redshift of these samples
described in Section~\ref{SS:theory}.  For example, for the lower mass LRG sample,
a mass of $3\times 10^{13}\hmsun$ gives a bias of $1.8$.  We compare
this number against the results in \cite{2005ApJ...621...22Z}, which gives a
clustering amplitude for this sample relative to $L_*$ of $1.85$.  
The highest richness maxBCG sample,
approaching $M\sim 6\times 10^{14}\hmsun$, has a bias of $5.5$ in this model.

We justify our neglect of the
stellar term by considering 
that a point mass at the origin gives a lensing signal $\ds = M_{point}/(\pi
R^2)$.  For the $L_*$ samples, typical stellar masses are a few $\times
10^{10} \hmsun$ and the minimum fit radius is $0.04$ \hmpc, which
implies that the signal from the stellar component at that radius is
subdominant compared to the observed signal of tens of
$hM_{\odot}/$pc$^2$.  For the LRGs, the stellar mass is higher by
factors of a few, but the minimum fit radius is $2.5$ times higher,
which again makes the stellar component subdominant on the scales used
for the fit.  Finally, for the maxBCG sample, 
given point masses of order $10^{12}\hmsun$ and at least $0.2$ \hmpc\
  for the minimum fit radius, it is difficult to
arrange for the stellar term to be more than 5 per cent of the
predicted signal from the NFW dark matter halo.

\subsubsection{Basic fits}

We begin by discussing the basic fits 1--3, which use minimum fit
radii of $0.2$, $0.5$, and $1$ \hmpc, without
explicitly accounting for failure to properly centroid BCGs.  We first
look for systematic effects due to BCG offsets by comparing the fit
results with different minimum radii.  The signature of BCG offsets
would be a lower concentration for a lower minimum fit radius; we see no
definitive sign of any such problem on the scales used here.  
For the case with minimum fit radius of $0.5$ \hmpc, which we adopt as
a compromise between 
minimizing the systematic error error due to this effect while
maximizing the statistical contraining power, the concentration at
$M_{200b}=10^{14} \hmsun$ is $c_0=4.6\pm 
0.7$, with a power-law scaling with mass of $-\beta=-0.13\pm 0.07$.  The mass
normalization at $N_{200}=20$ is $M_{200b}=(1.56\pm 0.12) \times 10^{14} \hmsun$,
with a scaling with richness of $1.15\pm 0.14$.  The $\chi^2$ of 266
for 282 degrees of freedom  
indicates that the fit is acceptable, as shown in 
Figs.~\ref{F:signal}--\ref{F:signalLRG}.  Our minimum maxBCG richness bin, at $\langle
N_{200}\rangle=12.4$, therefore has mean mass and concentration in this model of
$M_{200b}=0.9\times 10^{14} \hmsun$ and $c_{200b}=4.7$.  

The best-fit
masses for the LRG and lower 
luminosity samples are shown in Table~\ref{T:massfits}, and as we
anticipated, the LRGs are in a lower mass range that does
not overlap that of the maxBCG sample.  The spectroscopic LRGs with
$M_{^{0.0}r} \ge -22.6$ 
therefore predominantly trace group-scale halos, below the
cluster-scale masses of the maxBCG sample with $N_{200}\ge 12$.  We
also confirm our previous results \citep{2006MNRAS.368..715M} that, at
fixed $r$ band luminosity, isolated $L_*$ spirals have 
a lower mass than isolated $L_*$ ellipticals.
The signal
detections are higher significance than the $\sigma_M/M$ values would
indicate for L3: the error distributions determined using the
bootstrap-resampled datasets are non-Gaussian and
well-separated from zero.

\subsubsection{Fits with offsets}

We next consider the other sets of fits in Table~\ref{T:cfits}.  Fits
4--6 differ from 1--3 only in the inclusion of a prescription given in 
\cite{2007arXiv0709.1159J} to
correct for the centroiding problem in the maxBCG catalog (with no change
in the way the lower mass samples were handled).  
This prescription,
derived from mock catalogs, is described in detail in \cite{2007arXiv0709.1159J}.  In
brief, it has a richness-dependent fraction of misidentified BCGs
(from 30\% at low richness to 20\% at high richness), and those that
are misidentified have a Gaussian distribution of projected separation
from the true centroid, with a scale radius of $0.42$ \hmpc. 

The key point to
consider is that if the model used to account for centroiding problems
is correct, then the best-fit concentrations should be independent of
the minimum scale used.  We see that this is {\em not} the case: the
concentrations are significantly elevated compared to previous results
for minimum fit radii of $0.2$ and $0.5$ \hmpc, and differ from the
results with a minimum fit radius of $1$ \hmpc, for which (as one
expects) the offsetting prescription does not significantly change the
results given that the minimum fit radius is well outside the offset
scale.  Thus, we suggest that this particular offset procedure, which is
determined using mock catalogs in \cite{2007arXiv0709.1159J}, may in
fact overcompensate for the true level of BCG offsets.  This would not
be too surprising, given that it depends sensitively on the contents
of the mock catalogs, and is only applicable if they are a very true
representation of the real world.  At the high-mass end, the
comparison of maxBCG versus X-ray centroids in
\cite{2007ApJ...660..239K} can be used to evaluate the offset
procedure derived from mock catalogs.  For the derived 20 per cent
failure rate, with a Gaussian scale length of $0.42$ \hmpc\ for the
projected offset, we would expect 6.9 and 8.7 per cent of the clusters
to have centroids with projected offsets of $0.25$--$0.5$ and
$0.5$--$1$ \hmpc, respectively.  Out of the 87 clusters with X-ray
matches within $1$ \hmpc, figure 14 in \cite{2007ApJ...660..239K}
shows 11 (12.6 per cent) and 4 (4.6 per cent) in those two ranges of
projected offsets.  While these
numbers are not formally inconsistent with the expectations from
the mocks (e.g., for the $0.5$--$1$ \hmpc\ offset range, the one-tailed
$P(n\le 4)=0.12$), had they been used to derive an empirical model for
the offset, it would entail a smaller correction than that derived
from the mocks at the scales which are most relevant for our adopted
fit 2.  Since there may be uncertainty in the X-ray centroids for
disturbed or merging clusters, even the lower estimate in
\cite{2007ApJ...660..239K} may be an overestimate. 

 The fact that
\cite{2007arXiv0706.0727H} find similar levels of offsets for
photometric LRGs from the centroids of cluster X-ray distributions
does not contradict this conclusion: an examination of those results
suggests that they only apply for masses above $\sim 5\times
10^{14}\hmsun$ (after converting mass definitions), which essentially
corresponds to the top two maxBCG bins considered here.  
Furthermore, the mean redshift for that study is higher than here,
which implies different levels of photometric redshift errors and
therefore different levels of BCG misidentification.  Thus, it may not
be applicable in detail here even for the top two bins.  Finally, the
use of higher mass clusters at higher redshifts should give a higher
fraction of clusters that experienced recent mergers, so again we
expect an overestimate of the effect for the full maxBCG sample. The true
level of offsets in the four lower bins, which dominate the $c(M)$
fits, is poorly constrained from the real data.

\subsubsection{Fits with fixed power-law scaling}

Next, to investigate the correlation between $c_0$ and $\beta$, we
consider fits 7--12, which are the same as 1--6 except with a fixed
$\beta=0.1$ (the theoretical value).  While Table~\ref{T:cfits} gives
the formal correlation coefficient between these parameters from the
fit covariance matrix, it is instructive to explicitly fix $\beta=0.1$
to see the effect on $c_0$.  As shown, there is some 
degeneracy between $c_0$ and $\beta$, but a comparison of e.g. fits 2
and 8 suggests that this degeneracy is not very strong for the fit we
have selected as our main result.  Note that for fits 10--12, fixing
$\beta$ has somewhat ameliorated the discrepancy between the fits
using the offsetting procedure with different minimum scales.
Nonetheless, the trend towards increasing $c$ with decreasing minimum
scale is suggestive of possible overcompensation for the true level of
the problem in reality.

As stated previously, we choose fit 2 as our most robust result, but
the difference between this and the other fits suggests a systematic
error that is comparable in size to the statistical error.  Without
better knowledge of the true level of offsets of the maxBCG from the
cluster centroids (in the statistical sense), it is impossible to
reduce this systematic error.  However, for masses below a few $\times
10^{14} \hmsun$, where X-ray cluster data are difficult to obtain with
sufficient resolution at these redshifts, there is no clear, simple
way to observationally constrain these offsets at this time.

\subsubsection{Systematic tests of fitting procedure}

Next, we discuss a few fits not included in the table that were
designed to test for systematic errors.  \cite{2008MNRAS.387..536G}
find, using the Millenium simulation, that DM halos can
be more properly described using the Einasto profile than the
NFW profile, 
\begin{equation} 
\rho(r) =\rho_s e^{(-2/\alpha)[(r/r_s)^{\alpha}-1]}, 
\end{equation}
where $\alpha$ has a weak mass dependence with a value around 0.15. 
 Thus, fits to NFW profiles in simulations can lead to
different concentrations depending on the scale used for the fits.
While the differences between Einasto and NFW profiles are most
significant well within the scale radius, where we do not probe using
weak lensing, we nonetheless test that our results are insensitive to
the choice of Einasto versus NFW profiles.  When doing a fit that is
comparable to fit 2 but with NFW profiles replaced by Einasto
profiles, we find the best-fit masses to be preserved, and the
best-fit $c_0=4.5\pm 0.7$ and $\beta=0.12\pm 0.7$.  The changes from
fit 2 are well within the $1\sigma$ error, so we conclude that
possible errors in best-fit parameters due to the use of NFW rather than
Einasto profiles are insignificant.


We also perform fit 2 using three broader bins in $N_{200}$ instead
of the six narrow bins used for the main results.  We find that the
recovered $c(M)$ relation is virtually unchanged; the main parameter
that varies is $\gamma$ in the $M(N_{200})$ relation, which becomes
shallower by $1.5\sigma$.  This result is as expected from 
\cite{2005MNRAS.362.1451M}: the broad bins are most problematic at
the highest mass end, where they reduce the best-fit mass.  Since we
are not fitting the bins for masses individually, but rather are
fitting for a power-law relation, the exponent of this relation is
consequently reduced. Given the size of this shift, and the fact that
our  narrower bins used for the main analysis should contain mass distributions less than an
order of magnitude wide, we do not ascribe significant systematic
error in the concentration-mass relation to the default bin size.

Finally, because of potential centroiding systematics in the maxBCG lensing sample
that should not be present for the lower luminosity or LRG samples, we
performed the fits without the maxBCG samples entirely.  In that case,
we find that $\beta$ is quite poorly constrained, so we fix it to
$0.1$ and compare against fit 8 in Table~\ref{T:cfits}, which also has
$\beta=1$ and only differs in that it includes the maxBCG sample.  In
this case, we find the best-fit $c_0=4.7\pm 0.7$, entirely consistent
with the results in the table.  This result suggests that our choice
of minimum fit radius has minimized systematic error due to maxBCG
centroiding errors to be well within the statistical error.

\begin{table*}
\begin{center}
\caption{Results of fits to lensing signal for several different minimum
  fit radii for the maxBCG sample split into six richness bins,
  with and without the prescription 
  to account for BCG offsets described in section~\ref{SS:theory}.
  Fits 1--6 have $\beta$ (the power-law slope for the $c(M)$ relation)
  free; it is fixed to the theoretical value of $0.1$ in fits 7--12.\label{T:cfits}} 
\begin{tabular}{cccccccccc}
\hline\hline
Fit & $R_{min}$ & Offsets & $c_0$ & $\beta$ & Corr($c_0,\beta$) & $M_0$ & $\gamma$ & $\chi^2/\nu$ \\
 & \hmpc & & & & & $10^{14}\hmsun$ & \\
\hline
1 & $0.2$ & No & $4.2\pm 0.5$ & $0.16\pm 0.07$ & $-0.37$ & $1.56\pm 0.12$ & $1.20\pm 0.14$ & $314.8/336$ \\
2 & $0.5$ & No & $4.6\pm 0.7$ & $0.13\pm 0.07$ & $-0.63$ & $1.56\pm 0.12$ & $1.15\pm 0.14$ & $266.2/282$ \\
3 & $1.0$ & No & $4.0\pm 0.9$ & $0.18\pm 0.09$ & $-0.75$ & $1.53\pm 0.12$ & $1.23\pm 0.18$ & $215.3/240$ \\
\hline
4 & $0.2$ & Yes & $5.8\pm 0.5$ & $0.03\pm 0.07$ & $-0.44$ & $1.69\pm 0.13$ & $1.07\pm 0.15$ & $317.1/336$ \\
5 & $0.5$ & Yes & $5.8\pm 0.7$ & $0.03\pm 0.07$ & $-0.67$ & $1.69\pm 0.12$ & $1.08\pm 0.14$ & $270.4/282$ \\
6 & $1.0$ & Yes & $4.2\pm 0.9$ & $0.16\pm 0.09$ & $-0.75$ & $1.61\pm 0.12$ & $1.22\pm 0.18$ & $214.3/240$ \\
\hline
7 & $0.2$ & No & $4.3\pm 0.4$ & $0.1$ & - & $1.53\pm 0.11$ & $1.17\pm 0.13$ & $315.9/337$ \\
8 & $0.5$ & No & $4.8\pm 0.6$ & $0.1$ & - & $1.55\pm 0.11$ & $1.14\pm 0.13$ & $266.2/283$ \\
9 & $1.0$ & No & $4.6\pm 0.8$ & $0.1$ & - & $1.50\pm 0.11$ & $1.19\pm 0.17$ & $215.7/241$ \\
\hline
10 & $0.2$ & Yes & $5.5\pm 0.4$ & $0.1$ & - & $1.73\pm 0.12$ & $1.10\pm 0.14$ & $318.2/337$ \\
11 & $0.5$ & Yes & $5.3\pm 0.6$ & $0.1$ & - & $1.72\pm 0.11$ & $1.11\pm 0.13$ & $271.3/283$ \\
12 & $1.0$ & Yes & $4.7\pm 0.8$ & $0.1$ & - & $1.58\pm 0.11$ & $1.19\pm 0.17$ & $214.5/241$ \\
\hline
\end{tabular}
\end{center}
\end{table*}
\begin{table}
\begin{center}
\caption{Best-fit masses for LRGs and lower luminosity lens samples
  from the simultaneous fits to the lensing signal corresponding to
  fit 2 in Table~\ref{T:cfits}.  \label{T:massfits}} 
\begin{tabular}{lc}
\hline\hline
Sample & Mass $M_{200b}$ ($10^{12}\hmsun$) \\
\hline
LRG (fainter) & $30\pm 3$ \\
LRG (brighter) & $56\pm 5$ \\
Elliptical L3 & $1.0\pm 0.4$ \\
Elliptical L4 & $1.8\pm 0.4$ \\
Elliptical L5faint & $4.5\pm 0.8$ \\
Spiral L3 & $0.9\pm 0.3$ \\
Spiral L4 & $0.8\pm 0.3$ \\
\hline
\end{tabular}
\end{center}
\end{table}

\subsection{Concentration-mass relation}

\begin{figure}
\begin{center}
\includegraphics[width=3.2in,angle=0]{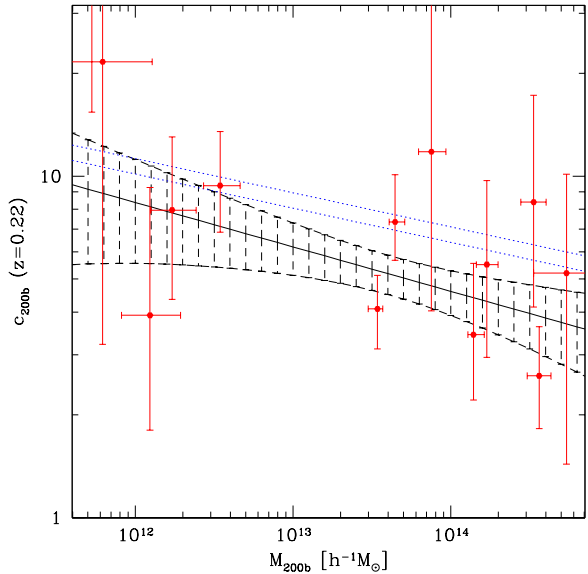}
\caption{\label{F:cmrelation}The best-fit $c(M)$ relation at $z=
  0.22$ with the
  $1\sigma$ allowed region indicated. The red points with errorbars
  show the best-fit masses and concentrations for each bin when we fit
them individually, without requiring a power-law $c(M)$ relation. The
  blue dotted lines show the predictions of
  \protect\cite{2007MNRAS.381.1450N} for our mass definition and 
  redshift, for the WMAP1 (higher) and WMAP3 (lower) cosmologies. The
  prediction for the WMAP5
cosmology falls in between the two and is not shown here.}
\end{center}
\end{figure}

In Fig.~\ref{F:cmrelation}, we show the best-fit $c(M)$ relation from
fit 2, with
a $1\sigma$ error region defined by the fits to fifty
bootstrap-resampled 
datasets.  As shown, the concentration-mass relation is best
constrained from $10^{13}$--$10^{14} \hmsun$, due to
the interplay between higher mass increasing the
lensing signal versus higher mass meaning a lower number density (and therefore
higher measurement error).  We
emphasize that this is the $c(M)$ relation at $z\sim 0.22$, so in the
simplest approximation of no mergers, the normalization at $z=0$
should be higher by about 20 per cent.  As shown, the range dominated
by the $\sim L_*$ samples ($10^{12} \hmsun$) yields a concentration of
$10\pm 3$, as expected theoretically.
At $6\times 10^{14} \hmsun$, the top end of the maxBCG sample, the
constraint is $4\pm 1$.  A constant concentration-mass relation is
just barely permitted at the $2\sigma$ level.  The red
points on the plot are the best-fit concentrations and masses for the
individual lens samples when we fit for $c$ and $M$ for each one
without requiring a power-law $c(M)$ relation.  The 
consistency with this power-law indicates that within the errorbars,
the $c(M)$ power-law is indeed a good fit to the data.

\subsection{Comparison against previous observations}

We can compare these results to our previous lensing results based solely 
on the LRG sample \citep{2006MNRAS.372..758M}. In that case we found
$c_{200b}=5.2 \pm 0.6$  at  
the pivot mass of $\sim 5\times 10^{13}\hmsun$, with 
weak constraints on the slope of the mass concentration relation
given the narrow mass range traced by LRG halos. This number is in good 
agreement with our fiducial value $c_{200b}=4.7 \pm 0.7$ at $10^{14}\hmsun$,
which gets increased by 10 per cent when going to the lower LRG
masses.  The LRG sample is one of the three samples used here and 
we follow essentially the same analysis, so the 
agreement is to some extent expected. 

Next we compare our results against the weak lensing determination of
$c(M)$ in \cite{2007arXiv0709.1159J}, which differs from ours in
several notable points: (1) we use the maxBCG sample to cover the range of
masses from $0.8$ to $6\times 10^{14} \hmsun$, whereas they use a
proprietary version of the catalog that extends roughly $1.5$ orders of
magnitude lower in mass; (2) we include several additional mass tracers 
extending the halo mass range a factor of $\sim 10$ lower than in 
\cite{2007arXiv0709.1159J} with very different selection criteria;
(3) we avoid
scales that are affected significantly by BCG centroiding problems,
rather than using a correction procedure derived from mock catalogs;
and (4) the photometric redshifts that they use to 
determine source redshifts and therefore normalize the lensing signal
suppress the lensing signal by $\sim 15$--$20$ per cent \citep{2008MNRAS.386..781M}.  

In their table 10, they show fit results for power-law relations
between mass and richness, and concentration and richness.  We
consider their result for $M_{180b}$, which should differ from our 
results with $M_{200b}$ by only a few per cent.  At $N_{200}=20$, they
find a best-fit mass of $1.2\times 10^{14}$ \hmsun.  This number is lower
than our result in Table~\ref{T:cfits} by 22 per cent; however, there
is a straightforward reason for this difference.  Using a large
spectroscopic training sample, we have recently shown that the
photometric redshift algorithm
used for sources in that work leads to a suppression of the
lensing signal for these lens redshifts of $\sim 20$\%
\citep{2008MNRAS.386..781M} and hence leads to about 30\% 
underestimation of mass.  They also find a steeper scaling with
richness, $1.3$ instead of $1.15$ as in our work.  This
difference can be explained by the fact that they use a much larger
range of richnesses for the fit, $N_{200}\ge 1$ instead of $\ge 12$ as
in our work.  It is apparent from their figure 11 that if one
restricts to $N_{200}\ge 12$, the best-fit power-law should be 
shallower than the result for their full richness range, roughly consistent
with our result.

We also compare against their results for concentration as a function
of richness. They find $c_{180b}=6.14$ at $M_{180b}=1.2\times 10^{14}\hMsun$, 
but as we argue above, their mass is underestimated, 
so this should really correspond to $M_{180b}=1.5\times
10^{14}\hMsun$.  
Rescaling, with $\beta=-0.1$, we find their value at $M_{200b}=10^{14}\hMsun$
is $c_{200b}=6.4\pm 0.3$.
The central value is in good agreement with the value we find when following their 
procedure of correcting the halo center misidentification with mock 
catalogs, $c_{200b}=5.8 \pm 0.7$. The statistical error is larger in our 
case because we do not use the small scale information. Note that because
the mocks are likely not to 
be a completely realistic description of the effect, \cite{2007arXiv0709.1159J}
attach a relatively large systematic error of 30\% on top of the relatively small 
statistical error they obtain. Instead, we trade statistical power 
for reduction of systematic error by 
using the fits from $0.5$\hmpc, where the effects of 
halo center misidentification are less severe, in which case we find $c_{200b}=4.6\pm 0.7$ for 
our standard fit with $\beta=0.1$ and no offsets, and very similar values also for the fits from 
scales above $1$\hmpc, either with or without offsets. We conclude that our result is relatively 
insensitive to the offsetting procedure, and while the results agree within the errors, our 
concentrations are significantly lower. 

Our results can be compared to those from other methods that are
used to determine the density profiles.  Recent cluster X-ray and 
strong-weak lensing studies have
found that the profile is consistent with the NFW model, but in many cases with 
a concentration that is higher than predicted by the concordance cosmology 
implied by WMAP
\citep{2006ApJ...640..691V,2007ApJ...664..123B,2007MNRAS.379..190C,2007MNRAS.379..209S}. 
While some previous work concluded that this result implies a higher normalization cosmology, this 
interpretation may be premature. There are many alternative explanations that need to be explored, 
such as the use of information from scales below 100\hkpc\ in clusters, where 
baryons have a significant contribution to the density profile and tend to steepen 
the profile, therefore increasing the best-fit concentration. In addition, there are significant
effects of triaxiality on the formation of arcs in the strong lensing regime. 
In the case of X-ray analysis, additional sources of pressure support may complicate the 
reconstruction based on hydrostatic equilibrium, since only the thermal pressure can be measured
directly, while other sources of pressure, such as turbulence, cosmic rays or magnetic fields, cannot. 
In some cases, the gas may not be relaxed at all, and the hydrostatic equilibrium assumption is invalid.  Another possibility are
selection effects, since the dispersion of the concentration is large
and 
correlates with X-rays or strong lensing selection such that only high
concentration clusters near the mass threshold are in the sample
\citep{2007A&A...473..715F}.  
It is important to compare the different tracers for consistency, and our analysis provides a complementary 
approach that can be compared against these more traditional analyses
only once these other effects are understood in more detail. 

\subsection{Comparison to simulations}\label{SS:comparison-theory}

Next, we compare our concentration parameter fits to theoretical expectations. 
We show theoretical predictions on top of the data in Figure \ref{F:cmrelation}, using the results
from the Millenium simulations  
in \cite{2007MNRAS.381.1450N} and \cite{2008MNRAS.387..536G}.
Using
a full sample, including both unrelaxed and relaxed halos, \cite{2007MNRAS.381.1450N} find
$c_{200c}=4.67[M_{200c}/(10^{14}\hmsun)]^{-0.11}$ (see also \cite{2007MNRAS.378...55M} and \cite{2007arXiv0709.3933H}).  First, we move these
$z=0$ results to our mean redshift of $0.22$, lowering the amplitude
by $1/1.22$.  Then, carefully converting both the
mass and the concentration to account for the different halo
definitions, we find the corresponding relation to be
$c_{200b}=7.1[M_{200b}/(10^{14}\hmsun)]^{-0.1}$.  This prediction is for the
  Millenium simulation cosmology with $\Omega_m=0.25$ and
  $\sigma_8=0.9$; if we convert to $\sigma_8=0.82$ \citep{2006JCAP...10..014S,2008arXiv0803.0547K} by assuming that $c
  \propto M_{\rm nl}^{0.1}$ \citep{2001MNRAS.321..559B}, then the predicted
  amplitude of this relation is reduced from $7.1$ to $6.7$. Finally, 
the results of
  \cite{2008MNRAS.387..536G,2008arXiv0804.2486D} suggest that the
concentrations in  \cite{2007MNRAS.381.1450N}  are too high by 10 per
  cent at the high masses where we have the most statistical power,
so this would bring the predicted value to 6.0 for the WMAP5 cosmology. 
Note that the scaling with mass no longer holds at the high mass end, where 
concentration becomes constant and is given roughly by $c_{200b} \sim 5-6$. 
Simulations predict that this occurs at masses comparable to or slightly higher than 
our highest mass bin, so we will continue to use the power-law scaling
with mass in our analysis. 
To be more quantitative we plot the
  predictions for WMAP1 and WMAP3 cosmologies at $z=0.22$ for the $M_{200b}$ mass
  definition on Fig.~\ref{F:cmrelation} together with the observational constraints.  As shown, the
  results for the lower normalization cosmology are $\sim 2\sigma$
  above our measured concentration at $10^{14}\hmsun$ (fits 2 or 8 in table \ref{T:cfits}).  

Typically the predicted profiles  are derived from N-body simulations
by fitting an NFW 
or Einasto profile to the 3d density distribution and averaging the obtained profile over all 
halos of a certain mass. 
There are many reasons why our observational procedure may differ from this. 
One is that we have both scatter in the mass-concentration relation and 
deviations from sphericity, both of which can
change the mean profile of 2-d $\Delta \Sigma$ when compared to the average density in 3-d.  
We also have scatter in the mass-richness relation, so that the assumption of a narrow 
mass distribution may be violated.  When
using the signal for central galaxies in the brightest luminosity bin 
in the simulations from \cite{2005MNRAS.362.1451M}, which incorporate
both scatter in the mass-luminosity relationship and in the
concentration-mass relationship, we found that the best-fit
concentration can be 20 per cent lower than if there is no scatter. However, 
the comparison to the expected concentration at the 
corresponding mass suggests that even without scatter, the
concentration fits can be biased 
by up to 20\%, possibly due to deviations of the average profile from NFW or Einasto, which 
show up differently in $\Delta \Sigma(R)$ than in the spherical radial profile $\rho(r)$. 
Another comparison in \cite{2006MNRAS.373.1159Y} also 
found that the concentration derived from $\Delta \Sigma(R)$
can be either an
underestimate (at low masses) relative to what is derived from the radial profile 
or an overestimate (at higher masses), but it is not 
clear how this result should be applied to our analysis since 
that analysis did not attempt to mimic our fitting procedure in detail. 

Another uncertainty arises from the predicted values for concentration in existing 
simulations. \cite{2004A&A...416..853D} find values that are about 10-20\% higher 
than the values used 
above. It is not clear how worrisome this is, given that it was derived from only a 
handful of preselected massive clusters. A more concerning issue is the difference between 
relaxed halos versus all halos when fitting for concentrations. When \cite{2007MNRAS.381.1450N} use 
relaxed clusters only when fitting for concentration, they find about 10\% higher values than for the full sample. 
Similarly, \cite{2007MNRAS.378...55M} find a large 
difference between the two, with relaxed clusters having typically 20\% higher concentrations at these masses. 
On the other hand, analysis of relaxed clusters in \cite{2008MNRAS.387..536G} can be compared to the full 
analysis presented in \cite{2007arXiv0709.3933H} and the latter gives only a 5\% reduction in the values of concentrations at the
halo masses around $10^{14}h^{-1}M_{\sun}$. 
Our sample consists of most or all halos above a certain mass threshold in a given volume, hence we should compare 
it to the full sample, which could bring the observed and predicted values into a better agreement. 

The above discussion suggests that there is some theoretical
uncertainty in the predicted values of concentrations, at a 
level of 20\%, that prevents us from making a more quantitative comparison to our results. 
This uncertainty could be reduced if exactly the same analysis 
used here on the real data is repeated on a large sample 
of simulated clusters, but doing so requires a large library of simulated galaxies and 
clusters from cosmological simulations with a volume comparable or larger to that used in the actual data analysis 
(of the order of ${\rm Gpc}^3$) yet with high mass resolution, for
which the next generation of simulations will be required, and 
is thus beyond the scope of this paper. 

\subsection{Implications for shear-shear lensing}

Our results also have implications for the theoretical interpretation of shear-shear lensing. 
The weak lensing power spectrum quantifies galaxy distortions produced by
lensing, which is sensitive to all matter in  
the universe, and as such it has long been advertised as being insensitive to
astrophysical uncertainties present in other tracers  
such as galaxies. However, this argument relies on the assumption that
the baryonic effects on the distribution of total matter  
can be understood. While earlier estimates found the effect to be small \citep{2004ApJ...616L..75Z,2004APh....22..211W}, 
recent work based on simulations with gas cooling 
finds that this assumption may be invalid \citep{2006ApJ...640L.119J,2008ApJ...672...19R}. 
They find that the baryons cause a significant redistribution of
matter within a halo, such that 
cooling and compression of baryons towards the center also makes the
dark matter more concentrated.  This redistribution can 
increase the concentration relative to the theoretical model
predictions by up to 40\% \citep{2008ApJ...672...19R,2008PhRvD..77d3507Z},  
and suggests that the matter profile is significantly 
redistributed well outside the inner region of the halo where gas has been transformed into stars. 

Observationally, we find no evidence for such an increase in concentration. We find that concentrations are 
at the lower end of the range predicted by simulations even for the low
normalization cosmology implied by WMAP3, and more so for the latest  
determinations of normalization and matter density \citep{2006JCAP...10..014S,2008arXiv0803.0547K}, which  
are already above the observations by $2 \sigma$. An increase in predicted concentrations 
by 40\% would make the discrepancy more than $5 \sigma$, which is ruled out by our analysis. If  halo center 
misidentification were considerably worse than estimated, then such a concentration enhancement would still be possible, 
but this would then be inconsistent with our analysis of halo
concentration determination as a function of inner radius (see table
\ref{T:cfits}),  
where we find no evidence of a systematic change in concentration with radius. It would also be 
inconsistent with our comparison between the analyses with and without
displacements, for which we find the difference to be  
within the statistical error if the analysis excludes information below
$0.5$\hmpc. Thus, on scales larger than this, 
effects due to baryon cooling are likely to be small, so that their
effect  on the existing shear-shear measurements 
can be neglected. 

In the future, the statistical error of shear-shear autocorrelation
measurements will be significantly reduced, so the baryonic effects
may therefore become 
significant, but at the same time, the measurements of the halo profiles
will improve as well. Thus, the effect can to some extent  
be corrected for by including the differences between the observed and
predicted galaxy-galaxy and cluster-galaxy lensing profiles in the analysis. One approach 
to do so is through the halo model analysis of the dark matter power
spectrum
\citep{2000MNRAS.318.1144P,2000ApJ...543..503M,2000MNRAS.318..203S,2001ApJ...546...20S},
which has been shown to give a good  
agreement with the simulations \citep{2008ApJ...672...19R}. A
necessary requirement, however, is that our
understanding of BCG  
halo center displacements improves, either through observations or via simulations. 

\section{Conclusions}\label{S:conclusions}

We used a large sample of 170~640 isolated spectroscopic galaxies, 38~236 groups traced by spectroscopic 
LRGs and 13~823 MaxBCG clusters  from SDSS, and applied a
weak lensing analysis 
to determine their average masses
and matter density profiles as a function of halo mass. 
We fit the lensing signal to an NFW or Einasto profile, 
excluding small scales 
to reduce the effects of 
baryons and misidentification of halo centers. The largest scale we use in the fits is 
3\hmpc\ for clusters, where the large scale structure contribution is
still small, but we account for it using 
a halo-halo term with a bias predicted by simulations.  For galaxies, the largest scale
we use is 500\hkpc, and for groups/LRGs, 1\hmpc.  

Fitting the lensing signal to an NFW profile, we find that the cluster 
concentration weakly decreases with mass,  
$c=c_0(M/M_0)^{-\beta}$, with $\beta=0.13\pm 0.07$ in good agreement with 
predictions from simulations.
The mean concentration at $M_{200b}=10^{14}\hmsun$ is $c_{200b}=4.6
\pm 0.7$ ($z=0.22$).  
This value should be compared to the predicted value  
$\sim 6$ for the best fit cosmological models \citep{2006JCAP...10..014S,2008arXiv0803.0547K}. 
The measured concentrations are below the predictions, although within $2\sigma$.
We find very little difference between NFW and Einasto profile in terms
of the measured concentration. 

While there appears to be a mild discrepancy between the predictions and observational constraints, 
there are significant uncertainties in the theoretical predictions that prevent us from  robustly 
concluding whether there is a problem.  
One task for the 
future is to repeat exactly the same analysis as done here on 
on a representative sample of halos from cosmological simulations 
covering the mass range of observed halos.
This should be possible in the near future
as a new generation of large volume and high mass resolution 
N-body simulations becomes available, thus allowing for
a more accurate calibration of the concentration-mass relation than is possible 
at the moment. 

However, to reduce the systematic uncertainty further we also need a better 
understanding of the displacement of BCGs from the halo center, which can be achieved 
either through observational studies \citep{2007arXiv0706.0727H} or through improved modeling in cosmological simulations \citep{2007arXiv0709.1159J}. 
In this paper we attempt to minimize it by using information outside the 
central region where we expect the effects from baryons in the central galaxy and from 
misestimation of the cluster center
to be small. We see no evidence of systematic contamination in the sense that we find consistent 
results with and without accounting for the halo center misestimation, but we cannot completely exclude the 
possibility that there are residual effects at a level comparable to or 
below the statistical error. Even if we increase the measured concentrations by this amount, they do not exceed
the predicted values, and thus we see no evidence of an enhancement in concentrations due to baryonic cooling predicted by 
some simulations \citep{2006ApJ...640L.119J,2008ApJ...672...19R}. This result bodes well for existing and future 
shear-shear weak lensing analyses, in that the baryonic effects are likely to be small and confined to small 
scales, and that by comparing theoretical and observed profiles as done here, these effects can be corrected for. 

\section*{Acknowledgments}

We thank the anonymous referee for useful comments on the submitted
version of this paper.  R.M. is supported by NASA
through Hubble Fellowship grant \#HST-HF-01199.02-A awarded by the
Space Telescope Science Institute, which is operated by the
Association of Universities for Research in Astronomy, Inc., for NASA, 
under contract NAS 5-26555.  U.S. is supported by the Packard Foundation, NSF 
CAREER-0132953 and Swiss National Foundation (grant number
200021-116696/1).   C.H. is supported by DoE DE-FG03-92-ER40701.

Funding for the SDSS and SDSS-II has been provided by the Alfred
P. Sloan Foundation, the Participating Institutions, the National
Science Foundation, the U.S. Department of Energy, the National
Aeronautics and Space Administration, the Japanese Monbukagakusho, the
Max Planck Society, and the Higher Education Funding Council for
England. The SDSS Web Site is http://www.sdss.org/. 

The SDSS is managed by the Astrophysical Research Consortium for the
Participating Institutions. The Participating Institutions are the
American Museum of Natural History, Astrophysical Institute Potsdam,
University of Basel, University of Cambridge, Case Western Reserve
University, University of Chicago, Drexel University, Fermilab, the
Institute for Advanced Study, the Japan Participation Group, Johns
Hopkins University, the Joint Institute for Nuclear Astrophysics, the
Kavli Institute for Particle Astrophysics and Cosmology, the Korean
Scientist Group, the Chinese Academy of Sciences (LAMOST), Los Alamos
National Laboratory, the Max-Planck-Institute for Astronomy (MPIA),
the Max-Planck-Institute for Astrophysics (MPA), New Mexico State
University, Ohio State University, University of Pittsburgh,
University of Portsmouth, Princeton University, the United States
Naval Observatory, and the University of Washington. 

\bibliography{apjmnemonic,cosmo,cosmo_preprints}


\end{document}